\documentclass{osa-article}

\journal{oe}

\usepackage{subfigure}
\articletype{Research Article}

\definecolor{Red}{rgb}{0,0,0}
\newcommand{\DB}[1]{{\color{Red} #1}}
\definecolor{Mycolor2}{rgb}{0,0,0}
\newcommand{\DA}[1]{{\color{Mycolor2} #1}}
\definecolor{Blue}{rgb}{0,0,0}
\newcommand{\Nadia}[1]{{\color{Blue} #1}}
\definecolor{Green}{rgb}{0,0,0}

\definecolor{Mycolor}{rgb}{0,0,0}
\newcommand{\LastNadia}[1]{{\color{Red} #1}}

\definecolor{Blue2}{rgb}{0,0,0}
\newcommand{\Hb}[1]{{\color{Blue2} #1}}

\begin{document}

\title{Supermode-based second harmonic generation in a nonlinear interferometer}

\author{David Barral$^{1,2,*}$, Virginia D’Auria$^3$, Florent Doutre$^3$, Tommaso Lunghi$^3$, Sébastien Tanzilli$^3$, Alicia
Petronela Rambu$^4$, Sorin Tascu$^4$, Juan Ariel Levenson$^1$, Nadia Belabas$^{1,\dagger}$, Kamel Bencheikh$^{1,\dagger}$}
\address{$^{1}$Centre de Nanosciences et de Nanotechnologies C2N, CNRS, Universit\'e Paris-Saclay, 10 boulevard Thomas Gobert, 91120 Palaiseau, France}
\address{$^{2}$Laboratoire Kastler Brossel, Sorbonne Universit\'e, CNRS, ENS-PSL Research University, Coll\`ege de France, 4 place Jussieu, F-75252 Paris, France}
\address{$^{3}$ Université Côte d’Azur, CNRS, Institut de Physique de Nice (INPHYNI), Nice, France}
\address{$^{4}$ \Hb{Research Center on Advanced Materials and Technologies, Department of Exact and Natural Science, Institute of Interdisciplinary Research, Alexandru Ioan Cuza University of Iasi, 700506 Iasi, Romania}}

\vspace{0cm}
\hspace{-0.4cm}
\authormark{$\dagger$} These authors contributed equally

\vspace{0.1cm}
\email{\authormark{*}david.barral@lkb.ens.fr} 

\begin{abstract}
We demonstrate supermode-based second harmonic generation in an integrated nonlinear interferometer made of linear and nonlinear directional couplers. We use a fully-fibered pump shaper to demonstrate second harmonic generation \DA{pumped by} the symmetric or anti-symmetric fundamental \DA{spatial} modes. The selection of the pumping mode and thus \Nadia{of} a specific SHG spectral profile is achieved through the selection of the fundamental wavelength \Nadia{and via a robust phase setting scheme. We use two methods: either post-selecting or actively setting the pumping mode. Such a modal phase matching \DA{paves the way for classical and quantum applications} of coupled nonlinear photonic circuits, where multimode excitation, encoding and detection are a route for multiplexing and scaling up light-processing.} 
\end{abstract}

\section{Introduction}

\begin{figure}[t]
\centering
{\subfigure{\includegraphics[width=0.4\textwidth]{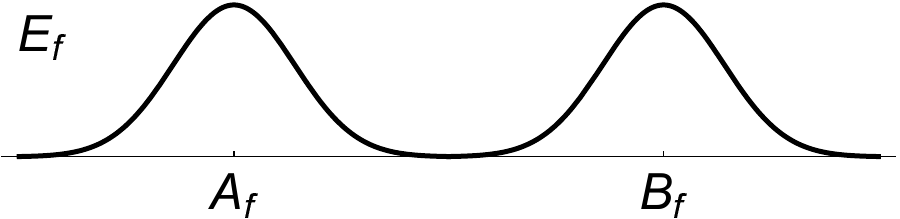}}} \hspace{1cm}
{\subfigure{\includegraphics[width=0.4\textwidth]{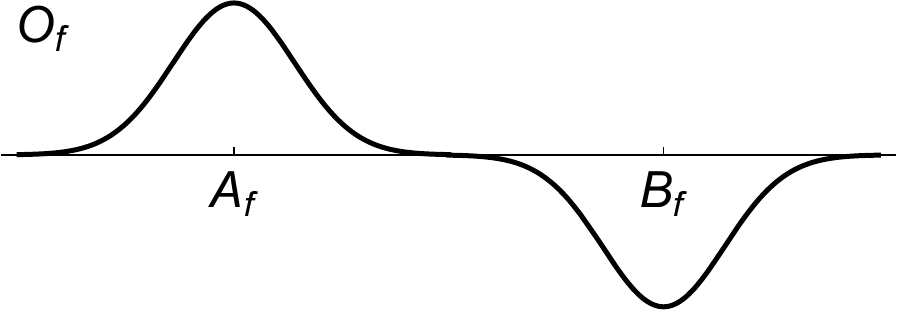}}} 
\caption{\label{F0} \DB{Even (left) and odd (right) fundamental supermodes of a directional coupler.}} 
\end{figure}

The nonlinear directional coupler (NDC) is a core device in integrated optics without analog in bulk optics: evanescent coupling and nonlinear interaction effects are intertwined instead of being sequential \Nadia{as it is the case for daisy-chained beamsplitters and active nonlinear elements}. The NDC has been widely studied in the classical regime \DA{for} ultrafast all-optical switching \cite{Jensen1982, Maier1983,Wright1988,Villeneuve1992,Assanto1993, Schiek1994, Schieck1996, Schiek1999, Hempelmann2002}. Recently, the $\chi^{(2)}$ NDC has found a flourishing field of application: quantum optics \cite{Perina2000}. Its key strengths in quantum information processing as a source of entangled photons and entangled field quadratures are actively explored \cite{Herec2003, Kruse2013, Kruse2015, Setzpfandt2016, Barral2017,Belsley2020}. Particularly, strong bipartite and four-partite continuous-variable entanglement \DA{are predicted in a second harmonic generation (SHG) configuration when pump light at the fundamental frequency is in propagation eigenmodes --i.e. in the spatial supermodes (see Figure \ref{F0}) \cite{Kapon1984}-- of the NDC \cite{Barral2018}. Such SHG schemes are especially interesting as they can harness widely available telecom domain sources to generate spatial entanglement at visible wavelengths compatible with atomic memories \cite{Hacker2016}. In addititon, they offer the possibility to generate in a controlled way multipartite entangled states, that are key resources for quantum technologies such as measurement-based quantum computing, multiuser quantum communications and quantum metrology \cite{Larsen2019, Asavanant2019, Roslund2013, Takeda2019, Guo2019}. In this paper we experimentally study SHG pumped by supermodes for the first time.
}

SHG in waveguides is commonly produced by quasi-phasematching \cite{Hum2007}. This method is based on periodical reset of wavevector phase mismatch to maintain a coherent buildup of the nonlinear interaction and it is conventionally obtained by periodic modulation of the nonlinear coefficient \DA{(periodic poling)} \cite{Armstrong1962, Somekh1972}. Ultra-high SHG conversion efficiencies in lithium niobate waveguides have been recently reported \cite{Kashiwazaki2020, Loncar2018, Boes2021}. \DA{In arrays of evanescently coupled waveguides, the poling period is usually fixed in order to fullfil the phasematching condition at  a given wavelength for the mode propagating in each waveguide. However, considering instead optical pump in spatial supermodes is a fruitful approach for multimode light processing in coupled waveguides and makes new and different phasematching conditions arise \cite{Barral2020, Barral2020b}.} The SHG is then generated efficiently at different wavelengths depending on the propagating supermode \DB{\cite{Kruse2015, Barral2019b}}. Supermode-based SHG has however remained elusive up to now due to the difficulty in exciting specific supermodes at the input of the array. In this paper we overcome those experimental issues for \Nadia{the NDC,} an array of two nonlinear waveguides, through a fully-fibered pump shaper and demonstrate supermode-based SHG through a specifically-designed integrated nonlinear interferometer made of linear (LDC) and nonlinear (NDC) directional couplers. \Hb{Note that the term nonlinear interferometer corresponds here to a linear interferometer with a phase changed due to the nonlinear effect \cite{Kitagawa1986}, in contrast with SU(1,1) nonlinear interferometers or similar schemes \cite{Chekhova2016, Luo2021}.}


The paper is organized as follows: in section \ref{Theory} we introduce the theoretical formalism that governs the SHG light in the nonlinear interferometer including the phase mismatch and the different pump configurations. In section \ref{Experiment} we detail the experimental setup. We exhibit experimental SHG generation in NDC via engineered pumping to access and harness the supermodes of the device in section \ref{Results}. Finally, the main results of this work are summarized in section \ref{Conclusions}.

\section{Theoretical model} \label{Theory}

\begin{figure}[t]
\centering\includegraphics[width=11cm]{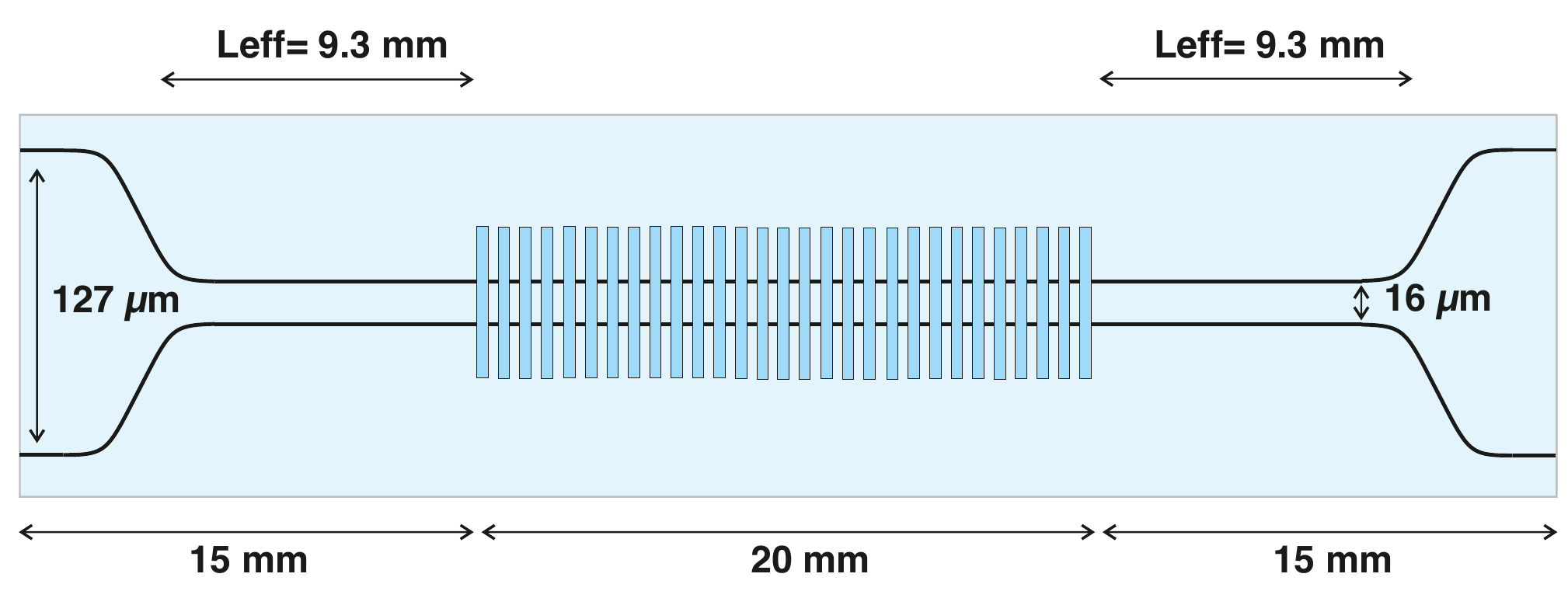}
\caption{\label{F1} Sketch of our nonlinear interferometer in lithium niobate. The input and output ports are \Nadia{127 $\mu$m apart to match the interfiber separation in a commercial V-groove array of fibers}. Waveguides are adiabatically coupled \Nadia{in 4.5 mm} into an 8.5 mm linear directional coupler (LDC) with 16 $\mu$m center to center separation. The evanescent coupling constant for the straight section is $C=0.25 \, mm^{-1}$. We compute the evanescently coupled region of the first 15 $mm$ of the chip are equivalent to a total effective coupling length of $L_{\text{eff}}=9.3$ mm (see text). $C L_{\text{eff}} \approx 3\pi/4$ and the LDC acts as a 3 dB beam splitter. The light then encounters a 20-mm long periodically poled region where SHG is produced in a NDC. The poling period is 16.4 $\mu m$ and the phase matching temperature 59.1 ºC. The device is symmetric with respect to the center.}
\end{figure}

\DB{Our device is composed by three sections of coupled waveguides produced by High Vacuum Proton Exchange (HiVacPE)  (Figure \ref{F1}):} a periodically poled lithium niobate waveguides-based NDC embedded between two LDCs. In the three sections, the energy of fundamental fields propagating in each waveguide is exchanged between the coupled waveguides through evanescent waves, whereas the interplay \Nadia{between} second harmonic fields is negligible for the considered propagation lengths due to their high confinement into the waveguides. We consider a single polarization mode in this model, since only TM fields can propagate in {Z-cut} \DB{HiVacPE} lithum niobate waveguides. The input-output LDCs are engineered to act as 3-dB beam splitters at the fundamental wavelength. The fundamental fields after passing a 3-dB LDC with two identical waveguides \Nadia{are}
\begin{align}\label{Tmat}
\begin{pmatrix}
A_{f}(z)\\B_{f}(z)
\end{pmatrix}=\frac{-1}{\sqrt{2}}
\begin{pmatrix}
1& -i\\
-i & 1
\end{pmatrix}
\begin{pmatrix}
A_{f}(0)\\ B_{f}(0)
\end{pmatrix},
\end{align}
where $A$ and $B$ are the slowly-varying amplitudes of fundamental ({\it f}) \DB{modes propagating in the upper (a) and lower (b) waveguides}, respectively, and $z$ is the coordinate along the direction of propagation. The input LDC allows to generate fields along specific axes of a Bloch sphere at the input of the NDC section \Nadia{at $z=15mm$}: 
i) pumping single waveguides \Nadia{i.e. $(1,0)^{T}$ or $(0,1)^{T}$ at $z=0$,} we obtain \DB{$(1,\pm i)^{T}/\sqrt{2}$}, where the superindex $T$ stands for transpose; 
ii) pumping the two waveguides with equal power and a relative phase $\Delta\theta$, i.e. \DB{$(1, e^{i\Delta\theta})^{T}/\sqrt{2}$} at $z=0$, yields \DB{$(\sqrt{1+\sin(\Delta\theta)}, \pm\sqrt{1-\sin(\Delta\theta)} )^{T}/\sqrt{2}$}, with $\pm$ respectively for $\Delta\theta \in \, [-\frac{\pi}{2}, \frac{\pi}{2}[$ and $\Delta\theta \in \, ]\frac{\pi}{2}, \frac{3\pi}{2}]$. Interestingly, \DA{this latter} configuration enables the excitation of the two propagation eigenmodes of the LDC: the even supermode  \DB{$E_f$, i.e. $(1,1)^{T}/\sqrt{2}$}, and the odd supermode \DB{$O_f$, i.e. $(1,-1)^{T}/\sqrt{2}$}, for respectively $\Delta\theta=0$ and $\pi$ \DA{(see Figure \ref{F0})} \cite{Kapon1984}.


The NDC section is made of two identical waveguides with $\chi^{(2)}$ nonlinear response due to periodic poling (PPLN). In each waveguide, the input fundamental field at frequency $\omega_{f}$ is up-converted into a second-harmonic field at frequency $\omega_{h}=2\omega_{f}$ (SHG). The physical processes involved in the NDC, \DA{i.e.} fundamental-fields evanescent coupling and nonlinear generation, are described \Nadia{in the perturbative regime} by the following system of equations \cite{Assanto1993}
\begin{align} \label{QS1}
\frac{d {A}_{f}}{d z}=&\, i C {B}_{f} +2 i g {A}_{h} {A}_{f}^{*}\,e^{i \Delta\beta z}, \quad \frac{d {A}_{h}}{d z}= i g {A}_{f}^{2}\,e^{-i \Delta\beta z}, \nonumber \\
\frac{d {B}_{f}}{d z}=&\, i C {A}_{f} +2 i g {B}_{h} {B}_{f}^{*}\,e^{i \Delta\beta z}, \quad \frac{d {B}_{h}}{d z}= i g {B}_{f}^{2}\,e^{-i \Delta\beta z},
\end{align}
where the subindex ({\it h}) refers to second harmonic fields, $g$ is the nonlinear constant, proportional to $\chi^{(2)}$ and \Nadia{to} the spatial overlap of the fundamental and harmonic fields in each waveguide, $C$ the linear coupling constant, and $\Delta\beta\equiv\beta(\lambda_{h})-2\beta(\lambda_{f})-2\pi/\Lambda$ is the wavevector phase mismatch with $\beta(\lambda)$ the propagation constant at wavelength $\lambda$ and $\Lambda$ the poling period. This system of equations is non integrable and thus does not present analytical solutions in general \cite{Bang1997}, \DA{although} analytical solutions in terms of Jacobi elliptic functions \DB{arise} under symmetric pumping conditions  \cite{Barral2019}.

Rewriting the Eqs (\ref{QS1}) for a NDC in terms of fundamental slowly-varying supermodes of a LDC, $E_{f}=(A_{f}+B_{f})e^{-iC z}/\sqrt{2}$ and $O_{f}=(B_{f}-A_{f})e^{iC z}/\sqrt{2}$, and linear combinations of harmonic modes $E_{h}=(A_{h}+B_{h})/\sqrt{2}$ and $O_{h}=(B_{h}-A_{h})/\sqrt{2}$, we have the following system of equations
\begin{align} \label{QS2}
\frac{d {E}_{f}}{d z}&= \sqrt{2} i g (E_{h} E_{f}^{*}\,e^{i (\Delta\beta-2C) z} + O_{h} O_{f}^{*}\,e^{i \Delta\beta z} ),\nonumber \\
\frac{d {O}_{f}}{d z}&= \sqrt{2} i g (E_{h} O_{f}^{*}\,e^{i (\Delta\beta+2C) z} + O_{h} E_{f}^{*}\,e^{i \Delta\beta z} ),\nonumber \\
\frac{d {E}_{h}}{d z}&=  i \frac{g}{\sqrt{2}}( E_{f}^{2}\,e^{-i (\Delta\beta-2C) z} + O_{f}^{2}\,e^{-i (\Delta\beta+2C) z}), 
\nonumber \\
\frac{d {O}_{h}}{d z}&=  \sqrt{2} i g E_{f} O_{f}\,e^{-i \Delta\beta z}.
\end{align}
\DB{$O_f$ and $E_f$ are fundamental supermodes, symmetric and antisymmetric, of the LDC region of the device, and can be also used as a suitable basis for the NDC region (Figure \ref{F0}). In contrast, and as the fields $A_h$ and $B_h$ are not coupled \DA{via evanescent fields}, $O_h$ and $E_h$ are just a convenient change of variable, they are not supermodes of either the NDC or LDC parts of the device.} \Hb{Note that in the case of non-negligible evanescent coupling of harmonic modes -- e.g. in the case of nanowaveguide arrays \cite{Boes2021}-- harmonic supermodes would appear modifying the dynamics of the system.}

\DB{The nonlinear and coupled Eqs (\ref{QS2}) do not have analytical solutions in general, but \DA{show} two phasematching conditions for the fundamental supermodes. Indeed, at the fundamental wavelength $\lambda_{f}=\lambda_{e}$, fulfilling the quasi-phase matching condition  $\Delta\beta(\lambda_{e})-2C=0$, the nonlinear interaction between the supermode field $E_{f}$ and the field $E_{h}$ builds up coherently and Eqs (\ref{QS2}) can be approximated by
\begin{align} \label{QS3}
\frac{d {E}_{f}}{d z}&= \sqrt{2} i g E_{h} E_{f}^{*}\,e^{i (\Delta\beta-2C) z} ,\nonumber \\
\frac{d {E}_{h}}{d z}&=  i \frac{g}{\sqrt{2}} E_{f}^{2}\,e^{-i (\Delta\beta-2C) z} , 
\end{align}
for wavelengths around $\lambda_{e}$. A similar condition is obtained for the fields $O_{f}$ and $E_{h}$ at the fundamental wavelength $\lambda_{f}=\lambda_{o}$ such that $\Delta\beta(\lambda_{o})+2C=0$ and
\begin{align} \label{QS4}
\frac{d {O}_{f}}{d z}&= \sqrt{2} i g E_{h} O_{f}^{*}\,e^{i (\Delta\beta+2C) z},\nonumber \\
\frac{d {E}_{h}}{d z}&=  i \frac{g}{\sqrt{2}} O_{f}^{2}\,e^{-i (\Delta\beta+2C) z}, 
\end{align}
for wavelengths around $\lambda_{o}$. The above approximations are correct as long as the coupling constant $C$ and the length of the NDC are large enough to avoid overlapping with the phasematching condition for the fundamental single modes $A_{f}$($B_{f}$), i.e. $\Delta\beta(\lambda_{single})=0$, around the wavelength $\lambda_{f}=\lambda_{\text{single}}$. Note that $O_{h}$ is coupled to both fundamental supermodes $E_{f}$ and $O_{f}$ in Eqs \eqref{QS2} with phasematching at $\Delta\beta(\lambda_{single})=0$. However, due to the spatial orthogonality of the fundamental supermodes\DA{,} it is not possible to excite both at the same time, so the generation of an antisymmetric linear combination of harmonic modes $O_{h}$ is not physically feasible.}

\DB{Eqs \eqref{QS3} and \eqref{QS4} are formally equivalent to the well-known equations of single mode SHG and have analytical solutions \cite{Armstrong1962}. In the low conversion regime the fundamental fields remain essentially constant over the interaction length, and the solutions for the harmonic supermodes when pumping the even and odd fundamental supermodes are respectively
\begin{align} \label{QS5}
E_{h}(z)= i \frac{g z}{\sqrt{2}} {E_{f}^{2}(0)} \, \text{sinc}(\frac{(\Delta\beta - 2C) z}{2}) e^{-i \frac{(\Delta\beta - 2C) z}{2}},\\  \label{QS6}
E_{h}(z)= i \frac{g z}{\sqrt{2}} {O_{f}^{2}(0)} \, \text{sinc}(\frac{(\Delta\beta + 2C) z}{2}) e^{-i \frac{(\Delta\beta + 2C) z}{2}}.
\end{align}
Thus, equal pump powers with phase difference $\Delta\phi=0$  ($\pi$) \cite{Note1} at the input of the NDC PPLN region at wavelengths $\lambda_{e}$ ($\lambda_{o}$) , i.e. excitation of the $E_{f}$ ($O_{f}$) supermode, generates efficiently a symmetric linear combination of harmonic fields $E_{h}$ around wavelengths $\lambda_{h}=\lambda_{e(o)}/2$. The mode $E_h$ builds up coherently when pumping at wavelengths $\lambda_{f}=\lambda_{e(o)}$ and the individual modes $A_h$ and $B_h$ of the individual single waveguides strive at $\lambda_{f}=\lambda_{\text{single}}$.}

\begin{figure}[t]
\centering\includegraphics[width=13.2cm]{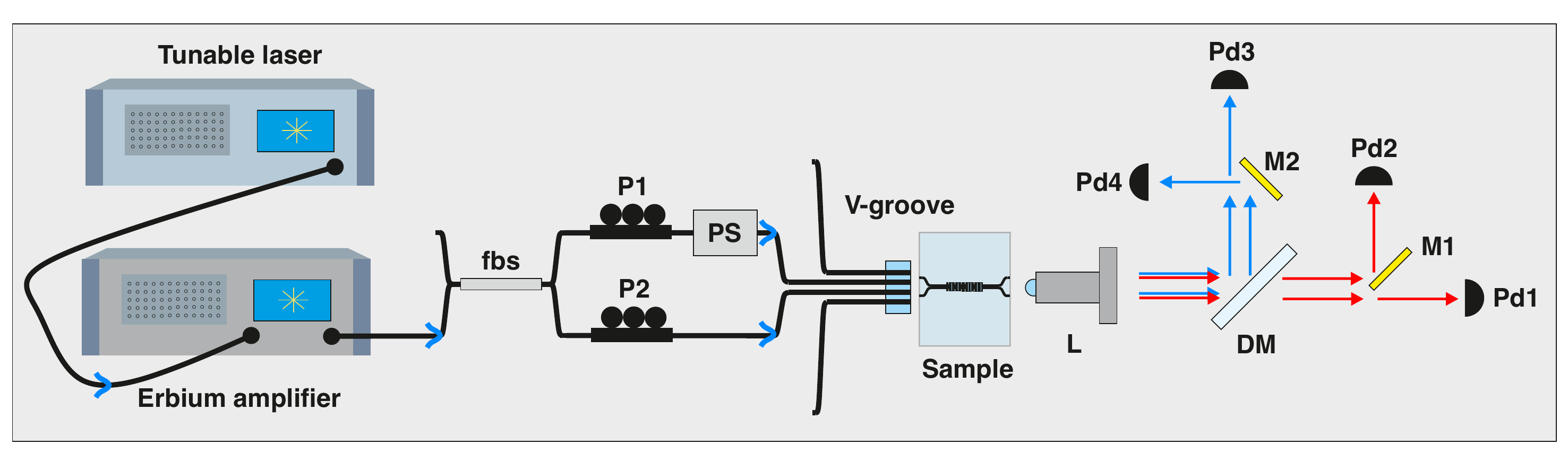}
\caption{\label{F2} Sketch of our experimental setup to measure supermode-based SHG. A Telecom fundamental field (blue) from a wavelength-tunable laser is amplified by an erbium amplifier. Then it is splitted in two by a fiber-beam splitter (fbs). Two fiber polarizers (P1 and P2) control the relative power injected in each waveguide that guide only TM-polarized modes. A fiber phase shifter (PS) controls the relative phase of the injected fields. The injection into the nonlinear interferometer is carried out via an AR-coated V-groove array of fibers. SHG fields (red) are produced on chip. The two spatial modes at fundamental and harmonic frequencies are collected with a lens (L) and frequency-demultiplexed with a dichroic mirror (DM). The fundamental (harmonic)  fields\DB{, drawn in blue (red)} are then directed to two photodiodes (Pd) sensitive to telecom (visible) frequencies with the help of a square mirror (M). Lenses in front of each photodiode are used to collect all the light (not shown). The four detectors (Pd1-Pd4) characterize completely the physical process in the nonlinear interferometer.}
\end{figure}

\section{Experiment} \label{Experiment}

In practice the nonlinear interferometer of Figure \ref{F1} has been implemented in high vaccum proton exchanged periodically poled Lithium Niobate \LastNadia{(HiVacPE PPLN) 6 µm wide-waveguides\cite{Rambu2018}}. As described in section \ref{Theory}, such a system is governed by the nonlinearity $g$ of the periodically poled region and the evanescent coupling $C$ between the coupled waveguides. To \DA{experimentally} determine the linear coupling constant $C$, we used arrays of 50-100 waveguides fabricated under the same conditions for different separations $d$ between waveguides. We obtained the following empiric dependance of the coupling constant as a function of the distance between the waveguides $C(d)=16.4 \,e^{-d/3.8}$ $mm^{-1}$ for $d \geq 13$ $\mu m$ \cite{Belabas2011}.
We designed the device to allow the use of commercial V-groove arrays of fibers with 127 $\mu m$ spacing between fibers for input and output. Our device has thus a $s$-shaped adiabatic region where the distance between the waveguides goes from 127 $\mu m$ to 16 $\mu m$ within a total length of 15 mm. The coupling strength for $d = 16$ $\mu m$ is $C=0.25$ $mm^{-1}$ {deduced from the empirical law}. Taking into account the coupling strength of the region where $d >16$ $\mu m$, the LDC could be approximated by an effective coupler with constant separation between waveguides and an effective length $L_\text{eff}=9.3$ $mm$. The corresponding coupling is such that $C L_\text{eff}\approx 3\pi/4$ \DA{and it is correctly} described by the transfer matrix of Eq. (\ref{Tmat}).

Our integrated nonlinear interferometer thus presents a NDC region embedded between two 3-dB LDCs (Figure \ref{F1}) where we obtain a 3-dB performance in the LDCs thanks to the coupling constant and bends engineering.   

The NDC section is a 20 mm periodically poled region. The poling period is $\Lambda=16.4$ $\mu m$. It has been chosen for SHG of $\lambda_{\text{single}} =1.56$ $\mu m$ radiation to produce $0.78$\,$\mu$m field at a temperature phase matching of 59.1\,$^{\circ}$C, stabilized within 0.1\,$^{\circ}$C with a resistive heater controlled through a PID controller (Thorlabs). The typical propagation losses of our waveguides are 0.2 dB cm$^{-1}$ at fundamental wavelength and 1 dB cm$^{-1}$ at SH wavelength. \LastNadia{The device (daisy-chained LDC+NDC+\DB{LDC}) total length is 5 cm.} To fully characterize our device, we need to determine the nonlinear strength $g$. This is done by measuring the SHG efficiency in a 6-$\mu$m-wide single waveguide of similar PPLN length, obtaining $\eta_{SHG}=42\%$ W$^{-1}$ cm$^{-2}$, which corresponds to $g\approx0.07$\,mm$^{-1}W^{-1/2}$. The refractive indices for fundamentals and harmonics modes are calculated \DB{with a finite element method simulation (COMSOL)} using the Sellmeier coefficients for bulk lithium niobate and a model of the refractive index of the \LastNadia{HiVacPE} waveguides \cite{Apetrei2017}. With these indices we calculate the $\Delta\beta$ we use in our {theoretical model}.

\begin{figure}[t]
{\subfigure{\includegraphics[width=0.47\textwidth]{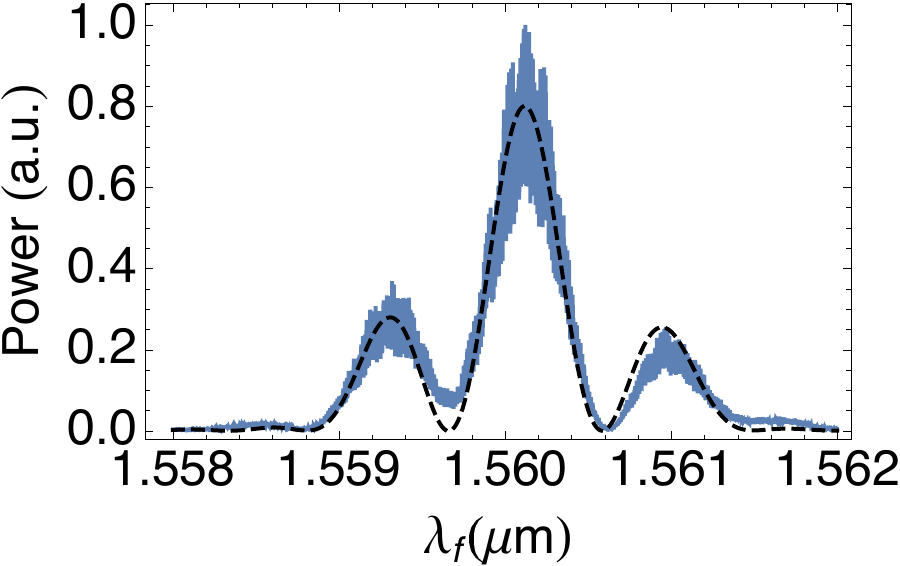}}} \hspace{0.2cm}
{\subfigure{\includegraphics[width=0.47\textwidth]{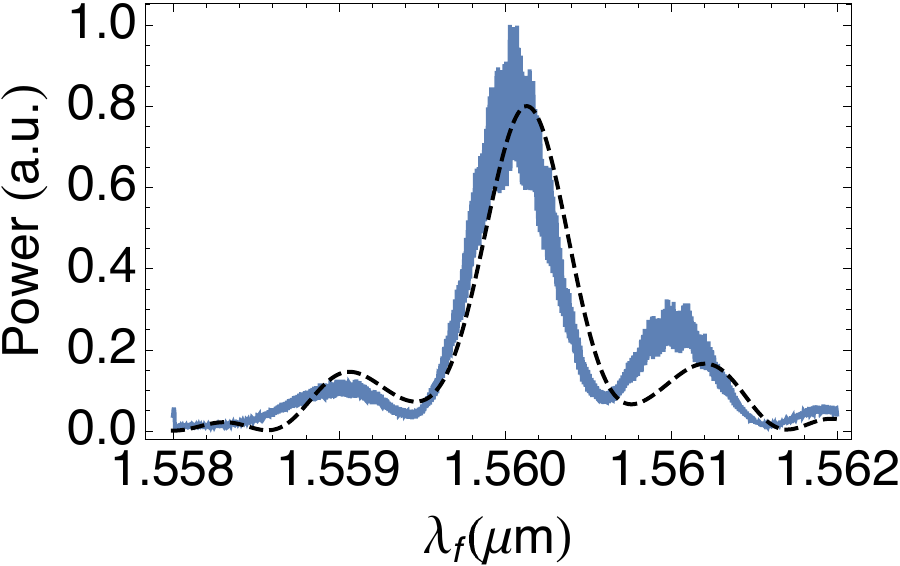}}} \\
{\subfigure{\includegraphics[width=0.47\textwidth]{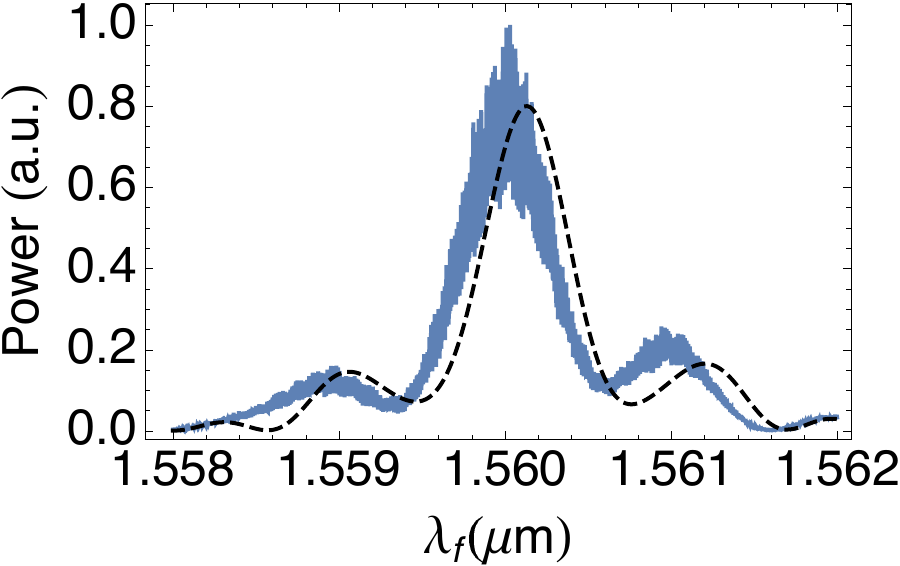}}}\hspace{0.2cm}
{\subfigure{\includegraphics[width=0.47\textwidth]{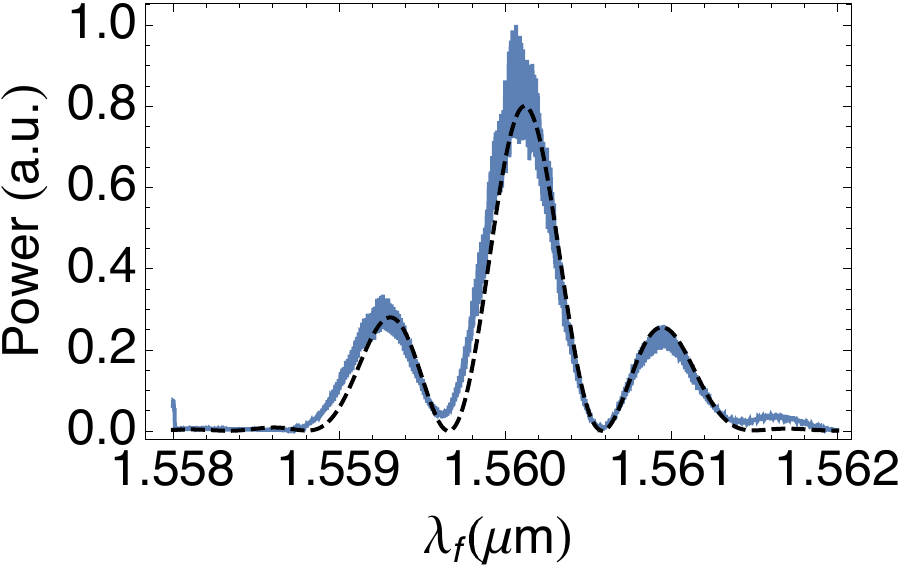}}}
\caption{\label{F3}Experimental single input-single output second harmonic generation in a nonlinear interferometer: $a\rightarrow a$ (upper left), $a\rightarrow b$ (upper right), $b\rightarrow a$ (lower left), and $b\rightarrow b$ (lower right). Dashed black lines are theoretical fits. \DB{Central wavelength at $\lambda_{f}\sim$1.5601 $\mu$m.} }
\end{figure}

The experimental arrangement used to study supermode-based SHG in a nonlinear interferometer is depicted in Figure \ref{F2}. It consists of cw tunable laser (TUNICS T100R) emitting 10 mW between 1490\,nm and 1620\,nm. The optical radiation is amplified with an Er-doped amplifier (KEOPSYS), generating up to 2 watts in the C-band. For the purpose of our investigation, we set the nominal laser power at 500\,$\mu$W (-3\,dBm) and amplify it up to 200 mW (23\,dBm). The telecom field (fundamental) is split in two paths by a 3dB fiber beam splitter and coupled into the chip by an AR-coated V-groove array of fibers with a pitch that \Nadia{matches} the waveguides spacing. Typical coupling efficiencies are estimated to be about $60\%$. The amplitude of each input is adjusted via 3-paddle polarizers that efficiently \Nadia{decrease} the amount of light coupled into the TM  guided mode of the \LastNadia{HiVacPE} waveguides without additional effect. The relative phase between each input is adjusted when necessary via a fiber phase shifter (IDIL FS20). The stability of the phase between inputs is better than $\lambda/20$ over few minutes, which {is the typical} duration of a full wavelength scan as depicted in Figures\,\ref{F3} to \ref{F5}. Both the polarization controllers and the phase shifter allow to fully describe the Bloch sphere and excite the NDC in all possible configurations. We collect the output signal from each waveguide with a collimating lens (THORLABS C-220 TMD-C) {having} a transmission $>85\%$ at the SHG wavelength and $>99\%$ at the fundamental wavelength. We separate the output fundamental fields from the harmonic fields using a dichroic mirror (THORLABS DMSP950). The two reflected fundamental beams are further separated by means of a gold mirror (M2 in Figure \ref{F2}) and  are focused on two InGaAs photodiodes (Newport 1811-FS) using two $f = 100$\,mm lenses. The two transmitted harmonic beams are also separated by means of a gold mirror (M1) and focused on two Si photodiodes (Newport 1801-FS) with two $f=150$ mm lenses. To improve the detection sensitivity of the generated SH beams, a 30\,\% depth intensity modulation is applied to the fundamental laser beam by driving its current with an RF signal. The modulated fundamental and harmonic intensities are then detected using SRS 830 {lock-in} amplifiers. The frequency of the modulation is fixed at 80 kHz to be compatible with the bandwidth of the lock-in amplifiers.  

\section{Results}\label{Results}

\subsection{Pumping a single waveguide}
\label{Single}

We start our analysis by exciting \DA{only} one of the inputs of the LDC. \DA{As described by Equation \eqref{Tmat}, due the linear coupling effect, the pump will be present at both inputs of the NDC. The relative powers and the relative phases will be defined by both the propagation length in the LDC and  the effective coupling constant. Accordingly, whatever the injected waveguide, the pumps arriving at the NDC inputs have equal amplitudes and a $\pi/2$ relative phase.} Figure \ref{F3} shows the SH spectra as the pump wavelength is tuned from 1.558 $\mu$m to 1.562 $\mu$m, exciting either waveguide $a$ or waveguide $b$. The SHG spectra are measured simultaneously at outputs $a$ and $b$ in both configurations. The two top plots correspond to the injection into one waveguide and the measurement of the SH spectra at the two output waveguides and the bottom plots are obtained when pumping the other waveguide. The apparent noise in the spectra is a Fabry-Perot periodic oscillation effect, not resolved in this scale, due to the {Fresnel} reflection at the input and output facets of the sample \DB{($\sim$14\% of reflectivity)}. The dashed black lines are theoretical results obtained from Eq.(\ref{QS1}) for $C_{0}=0.255 \pm 0.005$ mm$^{-1}$ and $g = 0.07$\,mm$^{-1}W^{-1/2}$, in excellent agreement with the measurements. This experimental value of $C_{0}$ is close to the one obtained from the empirical expression deduced from other experimental measurements, and in excellent agreement with the value of design thus demonstrating our ability to engineer the coupling constant. We note that our theoretical model fits with a high degree of accuracy the direct pumping configuration \DA{ input $a \rightarrow$ output $a$ (or $ b \rightarrow b$)}. However, there is a small wavelength mismatch for the indirect pumping configuration \DA{input $a \rightarrow$ output $b$ (or $ b \rightarrow a$)}. A small change in the phase mismatch $\Delta\beta$ used in our simulations can fit the indirect pumping traces at the cost of degrading those related to direct pumping. {Our theoretical model considers the temperature as homogeneous along the NDC. We hypothesize that our experimental result shows the impact of temperature \DA{inhomogeneity in the sample} on the SHG spectra and the sensitivity of SHG spectra measurement to the temperature stability \cite{Pelc2011,Santandrea2019A, Santandrea2019B}.}

We emphasize here the fact that the theoretical lines in Fig.(\ref{F3}) describing well the behaviour of the NDC are obtained using Eq.(\ref{QS1}), exciting the two inputs of the NDC with the same pump powers, $P_{A}=P_{B}$ ($\vert A_{f}\vert=\vert B_{f}\vert$) at $z=15$ mm, and with a $\pi/2$ relative phase. Any deviation from these conditions change drastically the shape of the fits. Thus, this demonstrates our control on the input fields at the NDC. Moreover, this shows our ability to control the LDC design and in particular to achieve a perfect 3\,dB integrated beamsplitter, with equal amplitude outputs and relative phase of $\pi/2$.

\DB{The spectra depicted in Fig.(\ref{F3}) exhibit  three peaks. The central peak at the fundamental wavelength $\lambda_{single}\sim$ 1.56 $\mu$m corresponds to the phasematching of single fundamental modes \DA{(i.e. $\Delta\beta=0$)}. The two side peaks appear as a consequence of the interplay between the phase mismatch at wavelengths $\lambda_{f}\neq \lambda_{single}$ and the beating of the fundamental mode amplitudes. \DB{Thus, the wavelengths of the side peaks are strongly dependent on the coupling coefficient.} Note that in this configuration there is no modal phasematching: the fundamental supermodes $E_{f}(O_{f})$ are never excited due to the $\pi/2$ relative phase between the fundamental single modes $A_{f}$ and $B_{f}$. Indeed, the fundamental field amplitudes oscillate between $(1, i)^{T}/\sqrt{2}$, $(0,1)^{T}$, $(1,-i)^{T}/\sqrt{2}$, and $(1,0)^{T}$ along propagation in the NDC due to the coupling. This effect produces a $z$-dependent phase mismatch between the fundamental modes and the generated harmonic modes. For wavelengths $\lambda_{f}\neq \lambda_{single}$ the energy flows forth and back from the fundamental modes to the harmonic modes at a rate given by the phase mismatch. Thus, for wavelengths very far from $\lambda_{single}$ the second harmonic generated is negligible. For the wavelengths where this beating is slow enough and that for the length of our NDC are in a cycle of harmonic generation --and not of parametric downconversion-- appear side harmonic peaks.

}


\subsection{Pumping both waveguides}

Now, we explore SHG of the NDC when exciting simultaneous both inputs of the LDC. The two fundamental fields are obtained after the amplifier using a fibered beamsplitter. With the LDC, they constitute a Mach-Zehner interferometer whose  two outputs feed the inputs of NDC \DA{(see Figure \ref{F2})}. Ideally, one would \Nadia{like} to equalize the \Nadia{NDC} input intensities and fix the relative phases to 0 or $\pi$ in order to excite either the symmetric supermode or the antisymmetric one. However, the LDC \Nadia{that is necessary to on-chip coupling} mixes the two inputs prior to the injection into the NDC according to Eq. (\ref{Tmat}). In the following, we introduce the two methods we explored and implemented to circumvent the issue of the LDC \Nadia{that demonstrate that this is an asset rather than a limitation. We first present how to extract the SHG spectra associated to a specific pumping mode via post-selection. For this scheme, we harness the simultaneous monitoring of the relative phase at the input of the NDC provided by the LDC. We further combine the SHG with a control of the phase at the input of the LDC which maps into a power balance control and enables supermode based generation at any fundamental wavelength.} 


\subsubsection{Length-difference phase}


\begin{figure}[t]
{\subfigure{\includegraphics[width=0.47\textwidth]{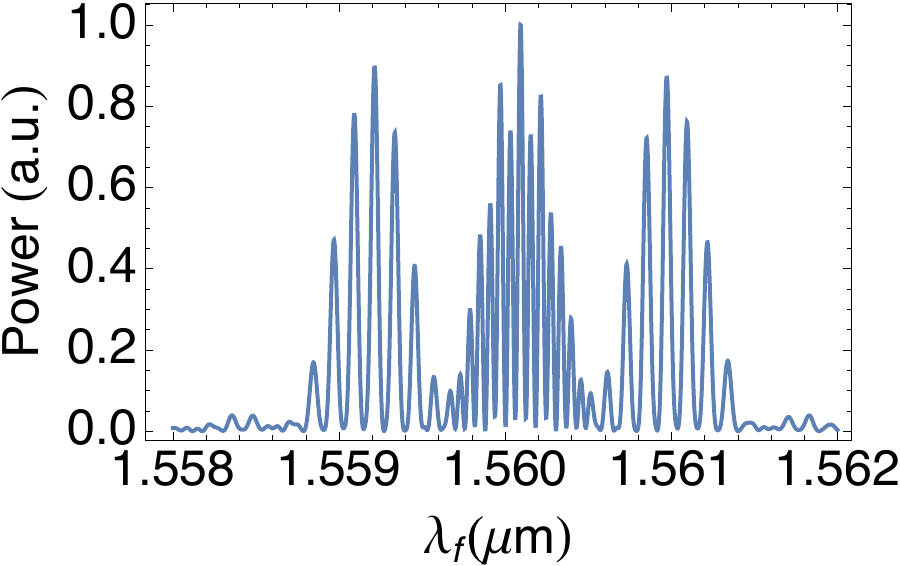}}} \hspace{0.2cm}
{\subfigure{\includegraphics[width=0.47\textwidth]{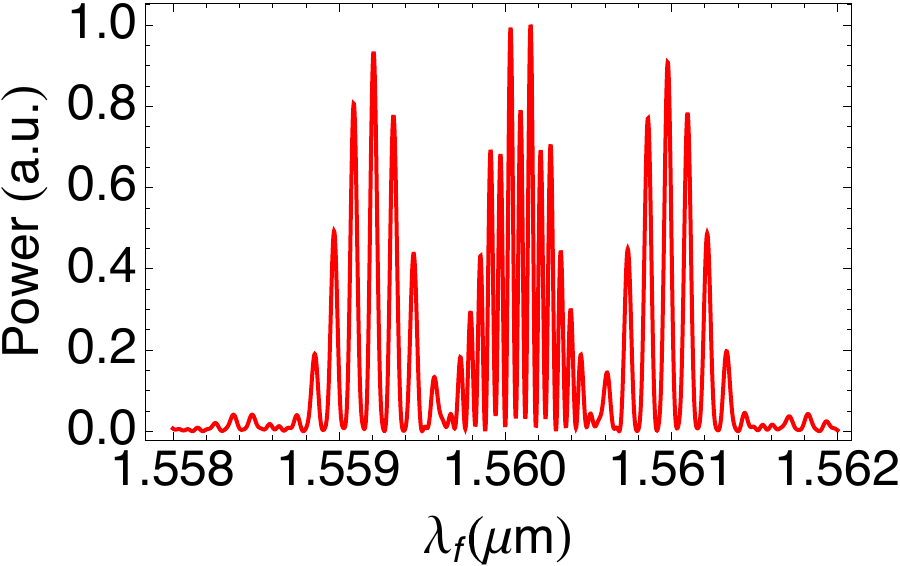}}} \\
{\subfigure{\includegraphics[width=0.47\textwidth]{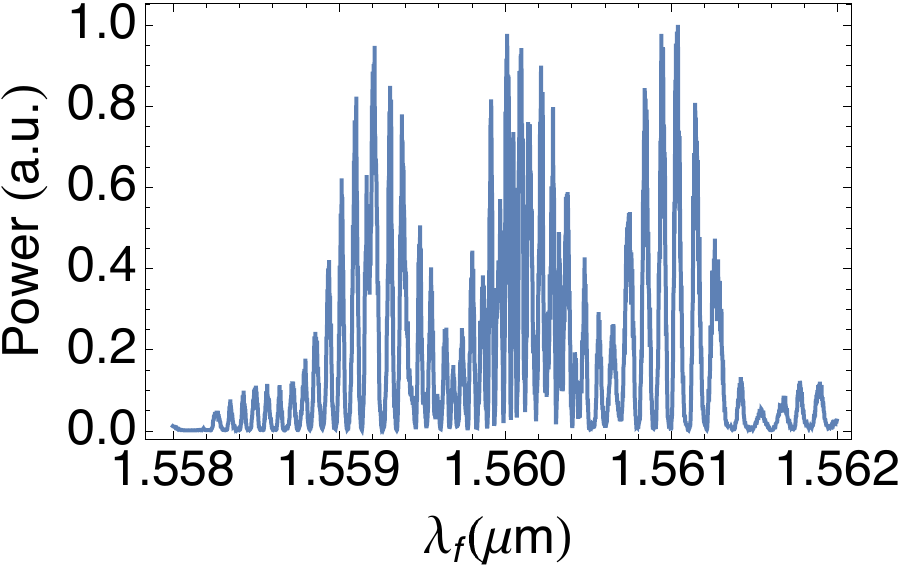}}}\hspace{0.2cm}
{\subfigure{\includegraphics[width=0.47\textwidth]{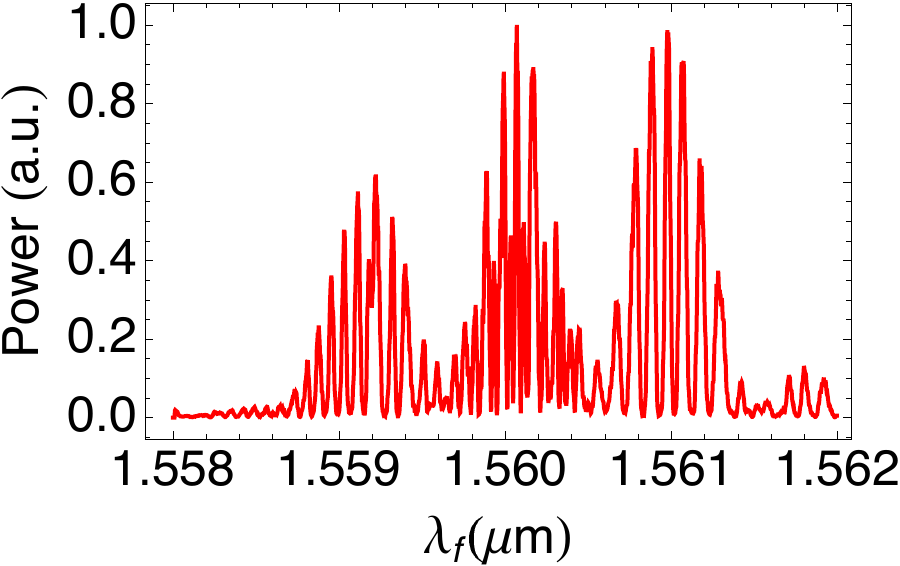}}}
\caption{\label{F4}Two-pump-input second harmonic generation in a nonlinear directional coupler with a $\lambda_{f}$-dependent relative phase: theoretical SHG (upper) in waveguides a (left) and b (right), and experimental counterpart (lower) for waveguides a (left) and b (right).}
\end{figure}
\begin{figure}[t]
\centering
{\subfigure{\includegraphics[width=0.65\textwidth]{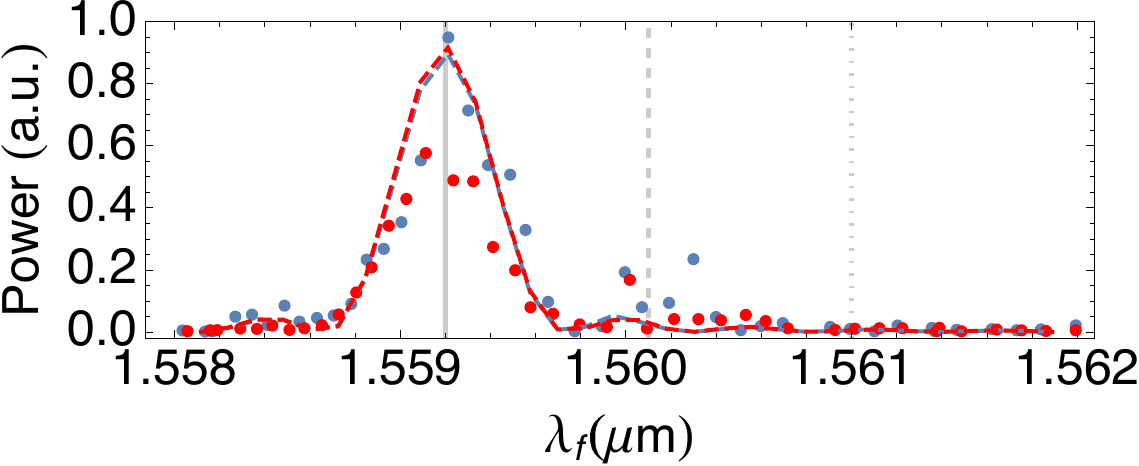}}} \\
{\subfigure{\includegraphics[width=0.65\textwidth]{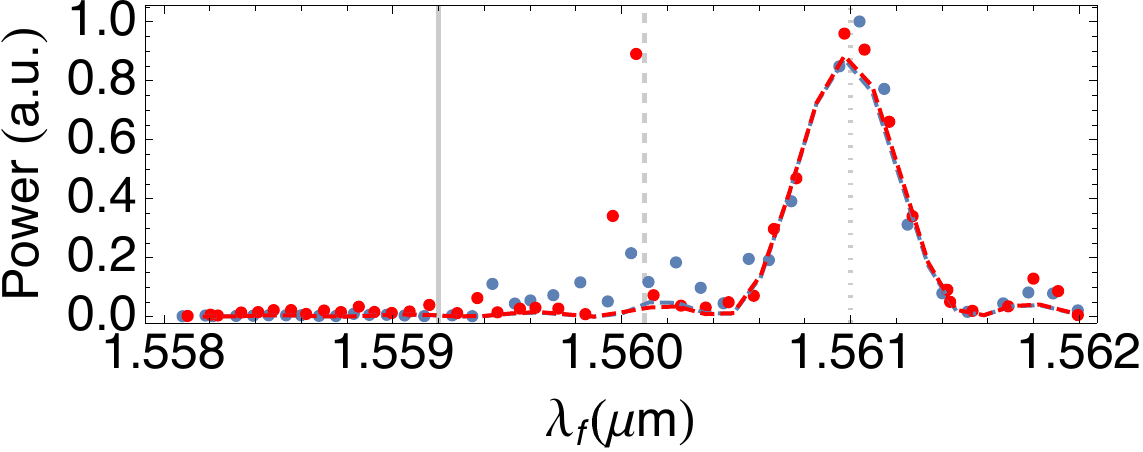}}}
\caption{\label{F5} Experimental supermode-based SHG sampled from the SHG of Figure \ref{F4} lower. SHG generated by the even fundamental supermode $E_{f}$ (upper) and by the odd fundamental supermode $O_{f}$ (lower). The generated harmonic modes $A_{h}$ and $B_{h}$ in blue and red, respectively. Dashed lines correspond to our theoretical model. Vertical lines stand for the phasematching fundamental wavelengths for even supermode ($\lambda_{f}=\lambda_{e}=1.5592$ $\mu m$, solid), individual modes ($\lambda_{f}=\lambda_{single}=1.5601$ $\mu m$, dashed) and odd supermode ($\lambda_{f}=\lambda_{o}=1.5610$ $\mu m$, dotted). \DA{The spurious points are due to phase jumps inherent to experimental instabilities in the interferometer.}} 
\end{figure}

We take \Nadia{advantage of the fact that the two arms of the Mach-Zehnder interferometer have, by construction, different lengths}. This is due to the \Nadia{small} difference in length of the \Nadia{similar} input fibers (Figure \ref{F2}). The fundamental fields experience a wavelength-dependent interference depending on the phase $\Delta\theta(\lambda_{f})=2\pi n_{c} \Delta L/\lambda_{f}$, where $n_{c}$ is the effective refractive index of the fundamental mode propagating in the fiber and $\Delta L$ is the length difference. Our \Hb{LDC} has been specifically designed such that $C L_\text{eff}=3\pi/4$. For such  $L_\text{eff}$ and equal input powers $P_{A}(0)=P_{B}(0)$, \DB{from Eq. \eqref{Tmat}} the fundamental powers and relative phase at the input of the PPLN NDC region are 
\begin{align}\nonumber
P_{A,B}(z=15 \text{mm})&=[1 \pm \sin(\Delta\theta(\lambda_{f}))] P_{A}(0),\\
\Delta\phi(z=15 \text{mm})=&\arg{[\frac{\cos(\Delta\theta(\lambda_{f}))}{1+\sin{(\Delta\theta(\lambda_{f}))}}]} =
\begin{cases} 0 \,\, \text{for} \,\, \Delta\theta \in [-\frac{\pi}{2}, \frac{\pi}{2}[ \\
\pi \,\, \text{for} \,\, \Delta\theta \in ]\frac{\pi}{2}, \frac{3\pi}{2}].
\end{cases} \label{pow}
\end{align}
\DB{Thus, according to Eq. (\ref{pow}), when tuning the wavelength $\lambda_{f}$, the relative phase of the fundamental fields reaching the NDC inputs is either 0 or $\pi$, exciting respectively the even $E_f$ or odd $O_f$ fundamental supermodes when $P_{A} = P_{B}$ at $z=15$ mm.} Conversely for wavelengths where $\Delta\theta(\lambda_{f})=\pi/2 \,(3\pi/2)$, only one of the two waveguides of the NDC, $a$ or $b$, is injected. In this way we can select the input {configuration} by only adjusting the fundamental wavelength. Depending on the choice of $\lambda_{f}$ the symmetric and antisymmetric fundamental supermodes are excited \DB{(Figure \ref{F0})} and act as pumps for SHG. The excellent control on fabrication parameters allows us to target and fabricate a device functioning at this convenient working point.

In order to demonstrate this performance, we first confirm that we pump equally both waveguides checking the ratio of fundamental powers at the output \Nadia{(Pd3 and Pd4 in Fig. \ref{F2})} when pumping first waveguide $a$ and then $b$. We then sweep \Nadia{the fundamental wavelength around 1.560 $\mu$m across a 4 $nm$-bandwidth} {with 1 $p$m resolution} and collect the SHG at each output. Figure \ref{F4} shows \DA{the theoretical predictions (upper) and the experimental results (lower).} \Hb{The discrepancy in the amplitudes between theory and experiment can be explained by a wavelength-dependent collection efficiency e.g. caused by an imperfect alignment of the output modes in the collecting lens. The number of oscillations showed is related to the length difference $\Delta L$ and it has been fitted manually in Figure \ref{F4} top-left and top-right to have a similar shape to the experimental spectra.}

Using the transmitted fundamental output in the upper waveguide $a$ as a reference, we follow the evolution of the relative phase as the wavelength is tuned. \Nadia{According to Eq.(\ref{pow}),} the maxima indeed correspond to $\Delta \theta = 0$ and thus $\Delta \phi = 0$ (even supermode), and the minima correspond to $\Delta \theta = \pi$ and thus $\Delta \phi = \pi$ (odd supermode). {We can thus select and separate the experimental data} in the SHG spectra corresponding to the maxima of the transmitted fundamental field ($\Delta \theta = 0$) from those corresponding to the minima ($\Delta \theta = \pi$). This is reported in Figure \ref{F5} in agreement with the theoretical expectations (dashed blue and red lines). The supermode-based SHGs are obtained for two fundamental pump phases with a $\pi$ difference. $\text{Sinc}^{2}$-shaped spectra appear centered at wavelengths equivalent to $\Delta\beta-2C=0$ (even) and $\Delta\beta+2C=0$ (odd) as expected from \DB{Eqs (\ref{QS5}-\ref{QS6}),} thus demonstrating supermode-based SHG. The spurious points in Figure \ref{F5} are due to phase jumps inherent to experimental instabilities in the interferometer. \Hb{This effect could be avoided by actively locking the phases of the fundamental fields.}

\subsubsection{Phase controlled}
\label{Phase}
In the previous section, the relative phase $\Delta \theta$ is tuned by scanning the laser wavelength. Hence, for each wavelength, we only have access to a single supermode related to the value of $\Delta \theta(\lambda_f)$. We have adopted an alternative strategy to access both supermodes and gain a new tuning parameter to balance pump power inputs. We have included in one arm of the interferometer a voltage-driven phase shifter achieving $2\pi$ phase shift with a 10 V triangular signal oscillating at 100 Hz. The full phase range becomes thus accessible for every wavelength. In the following we pump equally both waveguides and tune the phase \Nadia{$\Delta \theta$} from $-\pi/2$ to $3\pi/2$ thanks to the phase shifter.  

\begin{figure}[thb]
\centering
{\subfigure{\includegraphics[width=0.32\textwidth]{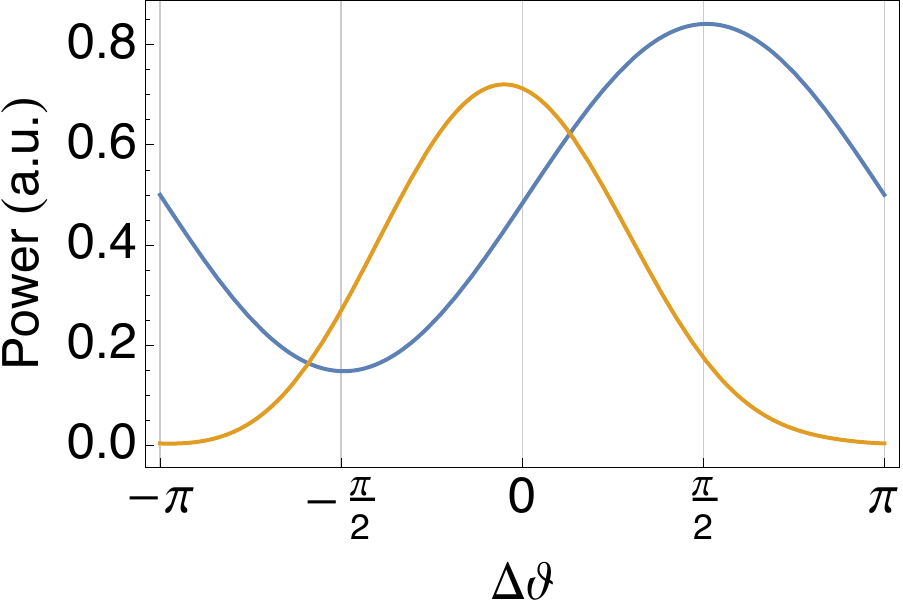}}} \hspace{0.1cm}
{\subfigure{\includegraphics[width=0.32\textwidth]{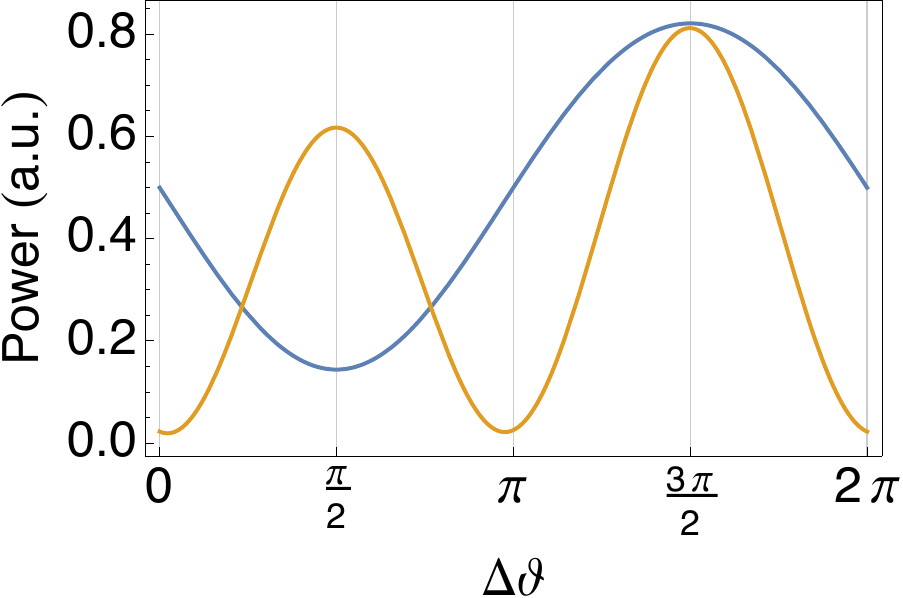}}} \hspace{0.1cm}
{\subfigure{\includegraphics[width=0.32\textwidth]{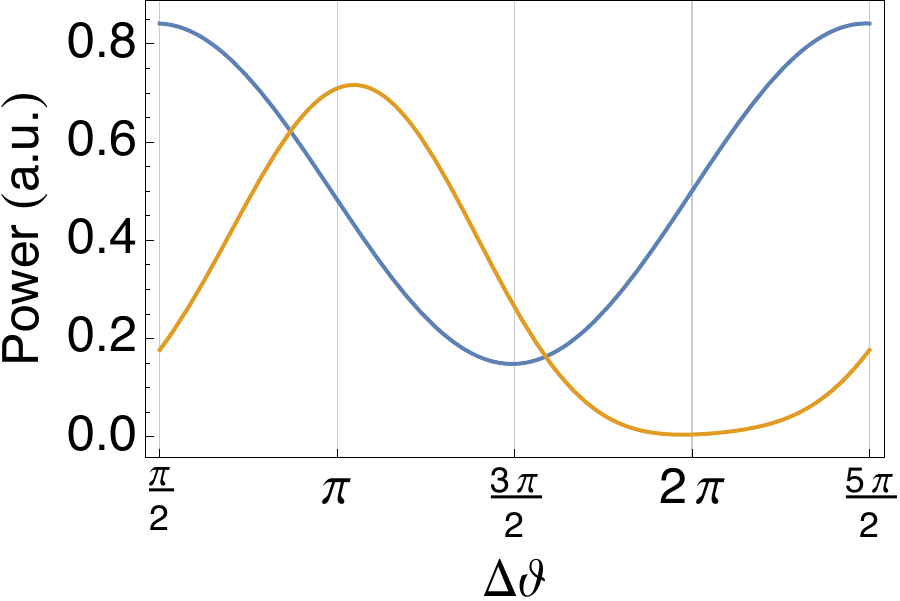}}}\\
{\subfigure{\includegraphics[width=0.32\textwidth]{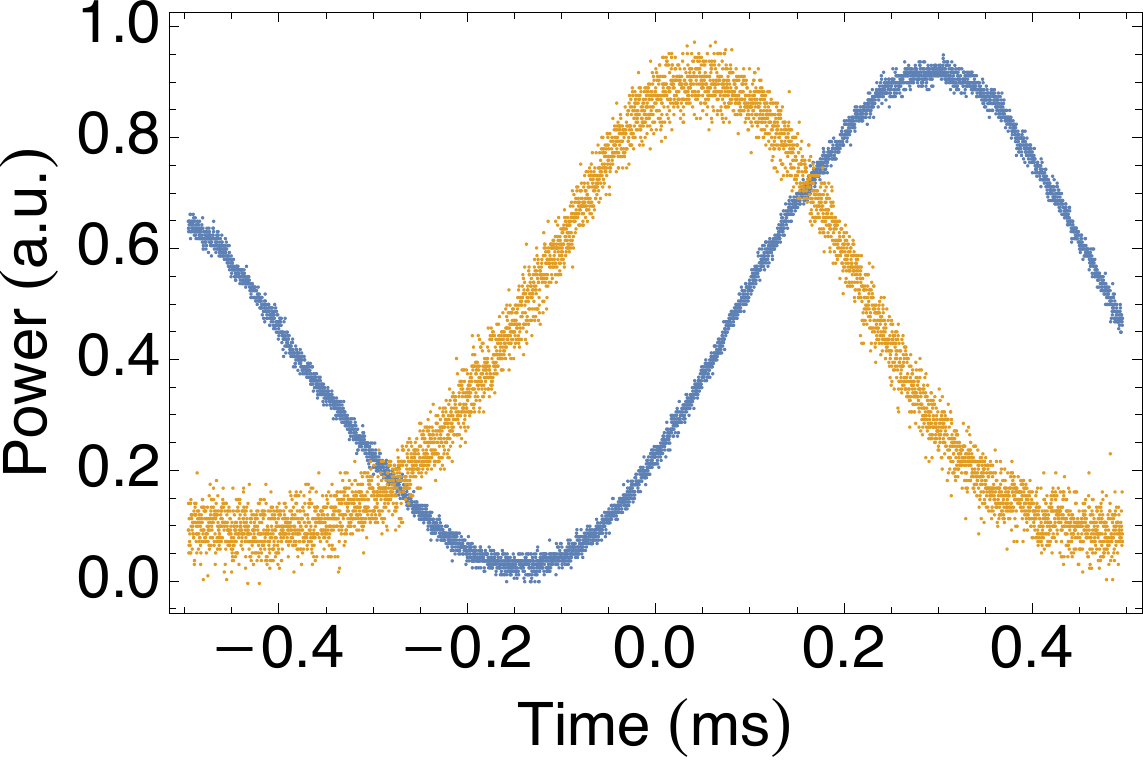}}} \hspace{0.1cm}
{\subfigure{\includegraphics[width=0.32\textwidth]{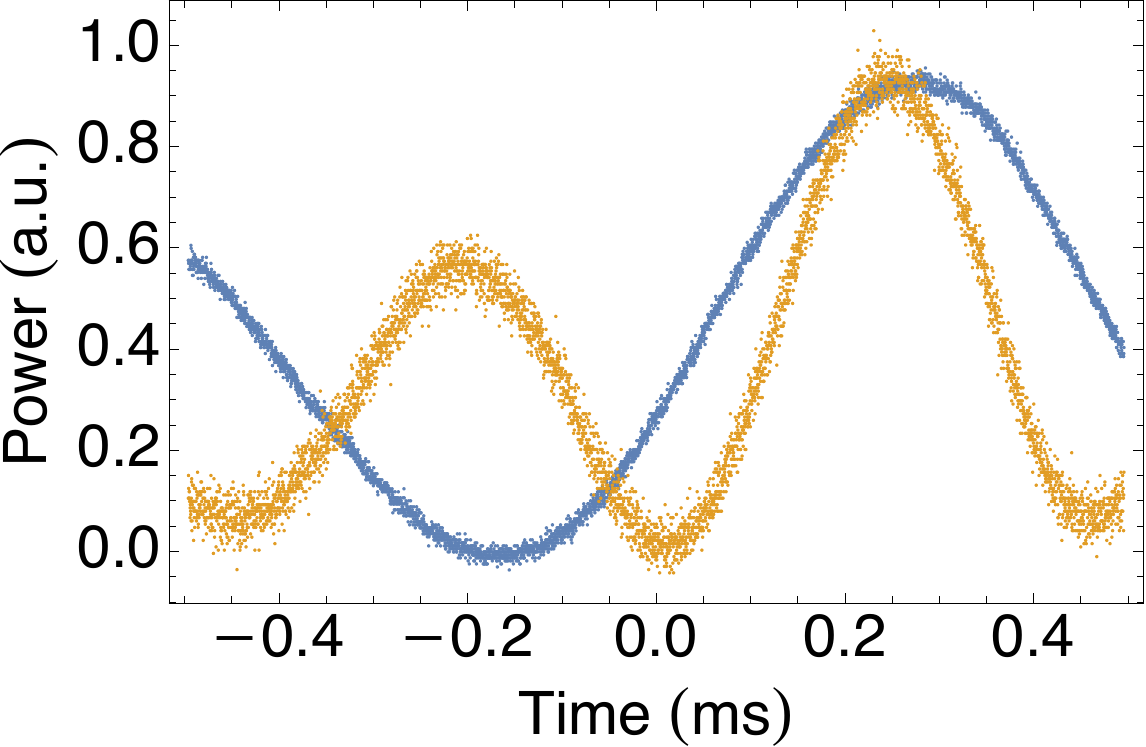}}} \hspace{0.1cm}
{\subfigure{\includegraphics[width=0.32\textwidth]{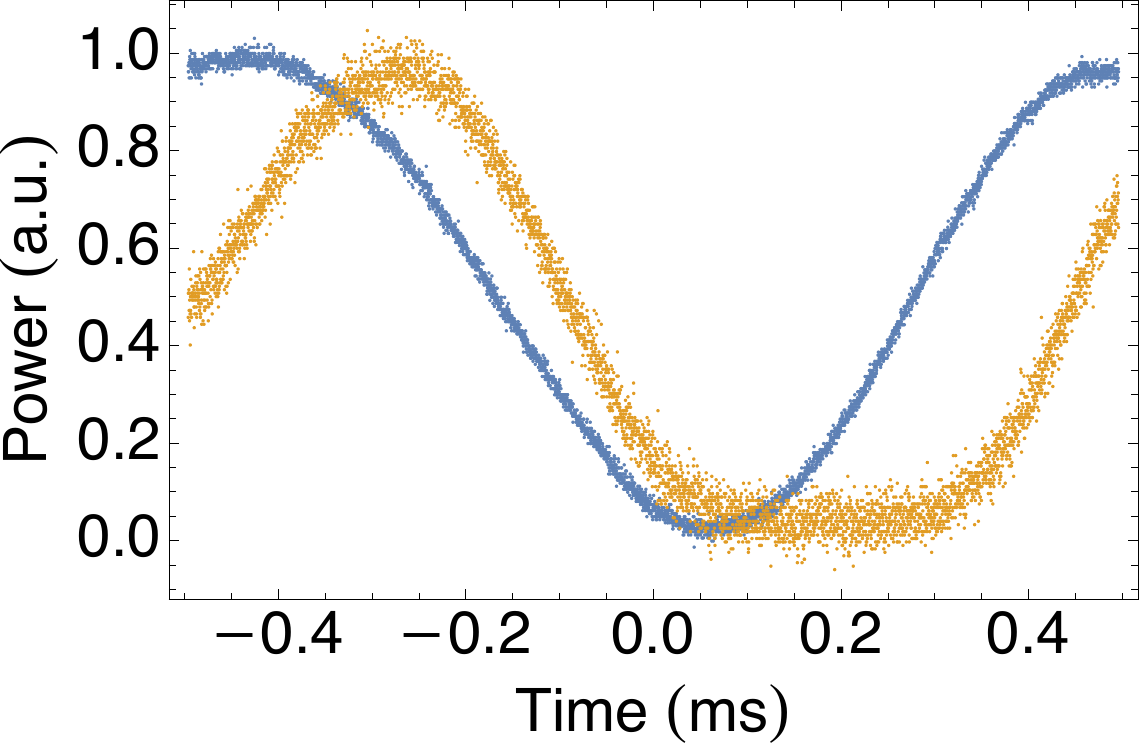}}}\\
\caption{\label{F6} Theoretical (upper) versus experimental (lower) SHG (yellow) for equal pump power in both waveguides and varying input phase $\Delta\theta$ at three phase-matching pump wavelengths $\lambda_{f}$ (see vertical lines in Fig. \ref{F5}). Left: even supermode-based SHG for $\lambda_{f}=\lambda_{e}=1.5592 \,\mu m$ ($\Delta\beta-2C=0$). Center: individual-mode based SHG for $\lambda_{f}=\lambda_{\text{single}}=1.5601\, \mu m$ ($\Delta\beta=0$). Right: odd supermode-based SHG for $\lambda_{f}=\lambda_{o}=1.5610 \,\mu m$ ($\Delta\beta+2C=0$). The fundamental field measured at the upper output $a$ which was modulated with a triangular signal at 480 Hz is used as a phase reference (blue). The windows on the x axis have been chosen to display conveniently the peaks as central.}
\end{figure}

The input phase $\Delta\theta$ is thus transformed into an adjustable pumping profile at the NDC region with $\Delta\theta$ as $P_{A,B}(L_{\text{eff}})=(1\pm \sin(\Delta\theta))P_{A}(0)$ with either 0 ($\Delta\theta \in [-\frac{\pi}{2}, \frac{\pi}{2}[$) or $\pi$ ($\Delta\theta \in ]\frac{\pi}{2}, \frac{3\pi}{2}]$) as relative phase \Nadia{$\Delta \phi$ at the input of the NDC}. Figure \ref{F6} shows the experimental results \DB{(lower)} and the related simulations \DB{(upper)} for a suitably chosen set of wavelengths. \Nadia{i) At $\lambda_{f}=1.5592$ $\mu m$ (left) we excite the even supermode at the NDC when $\Delta\theta=0$, producing supermode-based SHG (yellow). This wavelength is highlighted by the solid vertical lines in Fig. \ref{F5}. At $\Delta\theta=\pi$, however, no SHG is present as the odd supermode is not phase matched \Nadia{at $\lambda_{f}=1.5592$ $\mu m$}. ii) At the phase-matching wavelength of a single waveguide of 1.5601 $\mu m$ (dashed vertical lines in Fig. \ref{F5}), the supermodes are not phase matched, thus almost no SHG is produced both at 0 and $\pi$ phases. SHG is produced at $\pi/2$ and $3\pi/2$ since then individual modes are excited at the NDC. \DB{The physical process is here similar to that discussed in section \ref{Single}. The different height of the SH peaks is related to the effective interaction length seen by each individual fundamental mode when pumping one or the other input of the NDC.} iii) Finally, at $\lambda_{f}= 1.5610 \mu m$ (dotted vertical lines in Fig. \ref{F5}) we excite the odd supermode at the NDC when $\Delta\theta=\pi$, producing supermode-based SHG. At $\Delta\theta=2\pi$, however, no SHG is present as the even supermode is not phase matched at this wavelength. Note that the second LDC does not affect the generated second-harmonic fields as they are not evanescently coupled. \DB{This experiment demonstrates} our ability to produce supermode-based SHG at specific fundamental wavelength working points.}

This phase analysis at specific wavelength also sheds some light on the SHG observed in the case of the $\lambda_{f}$-phase dependence. Indeed Figure\,\ref{F6} also accounts for the double beating in the central peak of Figure\,\ref{F4}: around $\lambda_{\text{single}}$, the two peaks of SHG seen in the middle pannel of Figure\,\ref{F6} contribute whereas only a single peak contributes in the case of the side peaks around $\lambda_{e}$ and $\lambda_{o}$.

\section{Conclusions} \label{Conclusions}


We have experimentally demonstrated supermode-based second harmonic generation in an integrated nonlinear interferometer, made of linear and nonlinear directional couplers. We used a fibered and integrated pump shaper to engineer the pumping profile and thus exhibit SHG related to the symmetric and antisymmetric fundamental modes. More precisely, the selection of the pumping mode and thus of a specific SHG spectral profile is achieved through the selection of the fundamental wavelength that sets the pumping profile at the input of the non directional coupler. The experimental results match the theoretical predictions with an excellent agreement and without adjustable parameters. Indeed the governing parameters of the device - nonlinearity and coupling - have been characterized independently. 

Full characterization is done either by sampling second harmonic emission at the appropriate fundamental frequency working point or by changing the pump amplitude profile at a fixed fundamental wavelength. 
This latter scheme that actively sets the pumping mode incorporates a fibered phase shifter and a specifically designed linear directional coupler. This scheme maps the driving voltage of the phase shifter to the power balance at the input of the nonlinear direction coupler while enabling a switch between even-supermode or odd-supermode excitation. 

\DA{In addition to its interest in spatial multimode quantum optics, we stress that} this single shot supermode-based approach is fruitful to characterize a coupled nonlinear waveguide array device \cite{Lenzini2018}. We have combined on a single chip easy input and output coupling via V-groove fiber arrays and directional couplers and direct addressing of supermodes through a robust scheme that does produce reliably $0$ and $\pi$ dephasing at the input of the nonlinear directional coupler. The generated SH signal results from modal phasematching and is governed by coupling and nonlinearity. Such a multimode and modal approach is thus useful for integrated-optics characterization, especially as the widths, positions and height of the peaks in the SH spectra are degraded and displaced by inhomogenous poling, lack of accuracy for the coupling or inefficient nonlinearity \cite{Pelc2011,Santandrea2019A, Santandrea2019B}. From a quantum technologies point of view, these results are directly related to the harnessing of nonlinearities \cite{Mondain2019, Kashiwazaki2020} and supermodes \cite{Ra2020, Barral2020b, Gouzien2020} for quantum information and more specifically the generation and manipulation of advanced entangled states. Indeed, the NDC is the simplest of important devices for quantum information to generate robust entangled multicolor states in the discrete and continuous variable domain \cite{Barral2020}. Our fiber and integrated excitation scheme to adress the NDC is thus an interesting addition to explore with versatility the supermode-based and individual mode-based entanglement.

\section*{Acknowledgements} This work was supported by the Agence Nationale de la Recherche through the INQCA project (Grants No. PN-II-ID-JRP-RO-FR-2014-0013 and No. ANR-14-CE26-0038), the Paris Ile-de-France region in the framework of DIM SIRTEQ through the project OSTINATO, and the Investissements d'Avenir program (Labex NanoSaclay, reference ANR-10-LABX-0035).

\section*{Disclosures} 
The authors declare no conflicts of interest.

\section*{Data availability}
Data underlying the results presented in this paper may be obtained from the authors upon reasonable request.

\end{document}